# Current rectification via Photosystem I monolayers induced by their orientation on hydrophilic self-assembled monolayers on titanium nitride*

Jonathan Rojas, Zhe Wang, Feng Liu, Jerry A. Fereiro, Domenikos Chryssikos, Thomas Dittrich, Dario Leister, David Cahen and Marc Tornow

*Abstract*— Photosystem I (PSI) is a photosynthetic protein which evolved to efficiently transfer electrons through the thylakoid membrane. This remarkable process attracted the attention of the biomolecular electronics community, which aims to study and understand the underlying electronic transport through these proteins by contacting ensembles of PSI with solid-state metallic contacts. This paper extends published work of immobilizing monolayers of PSI with a specific orientation, by using organophosphonate self-assembled molecules with hydrophilic heads on ultra-flat titanium nitride. Electrical measurements carried out with eutectic GaIn top contacts showed current rectification ratios of up to ~200. The previously proposed rectification mechanism, relying on the protein's internal electric dipole, was inquired by measuring shifts in the work function. Our straightforward bottom-up fabrication method may allow for further experimental studies on PSI molecules, such as embedding them in solid-state, transparent top contact schemes for optoelectronic measurements.

## I. Introduction

Because of the technological advances in nanoscience research and the semiconductor industry, electronic devices at the (bio-) molecular level are now becoming an experimental reality [1-3]. This effort is stimulated by the existence of biomolecules, engineered by evolutionary mechanisms, which resulted in molecular complexes that, among other vital transformations, effectively move electric charges in space. One of the paramount examples of nature's design abilities is photosynthesis. This enzymatic process takes water as an analyte, converts it into molecular oxygen, and generates the biological energy currency, ATP and a reducing agent, NAD(P)H, via absorption of photons and transfer of electrons.

One of the proteins responsible for green plant and bacterial photosynthesis is the multi-subunit protein complex called photosystem I (PSI). It is a transmembrane protein comprising 95 light-sensitive chlorophyll molecules carefully organized as an antenna to feed photons to a chlorophyll dimer, the P700 pair (absorption maximum at 700 nm), where excitons are generated. After the exciton splits, the electron goes through a chain of molecules to its end-station, the iron-sulfide (FeS) acceptor [4, 5]. PSI is usually studied in its trimeric form because of the relatively high stability of the latter, which contrasts with the stability of its monomeric form [6]. The PSI trimer has a cylindrical shape of around ~20 nm in diameter and ~9 nanometers in height. Figure 1 a) illustrates a PSI monomer with the electron transfer chain: the P700 special chlorophyll pair is located near the lumen side and the final FeS acceptor is on the stromal side at the top of the protein. Extensive investigations suggested that PSI has a quantum efficiency of almost 1 [7] and an electron transfer time of around 16 nanoseconds, across the height of the protein (9 nm), in the dry state [8]. These astonishing parameters have motivated recent inquiries about the possibility of effectively harnessing PSI to power solar cells [9]. Another striking feature of PSI's electronic properties is its capability to rectify electrical current [6, 10]. Pioneering results on PSI individual reaction centers with scanning tunnelling microscopy (STM) established that diode-like behavior could emerge depending on the orientation of the electron transfer path with respect to the anchoring surface [11]. In more recent research, the PSI internal electric dipole, whose direction is from the stromal to the lumen side, contrary to the natural electron flow chain, has been proposed to be responsible for the rectification properties [10]. A potential application has been reported in the work of Qiu and Chiechi, who built logic circuits (from AND and OR gates) using perfectly oriented PSI monolayers as diodes and randomly oriented PSI monolayers as resistors [12].

One common strategy for immobilizing PSI molecules on a bottom substrate is using self-assembled monolayers (SAM) as linkers. A common substrate material is gold; therefore, thiols with amino or carboxyl ending groups are the molecules of choice for the linker SAM. PSI trimers, being transmembrane proteins, have hydrophilic amino acid groups at the lumen and stromal sides. Hence, they interact via hydrogen bonding with the polar groups of the linker SAM [6]. Different methods have been previously adopted to control the orientation of PS1 on metallic contacts [13]. For example, some residual groups can be mutated with genetic engineering, and cysteine groups can be deterministically positioned on the lumen and/or stromal sides of the protein [8, 14, 15]. Another study successfully oriented PSI using monolayers of C60 cages with carboxylic acid ending groups [12, 16]. All of them,

* Research was funded by the Deutsche Forschungsgemeinschaft (TO266/10-1) and EXC 2089 [excellence cluster e-conversion]

Jonathan Rojas (jonathan.rojas@tum.de), Zhe Wang, Domenikos Chryssikos and Marc Tornow (tornow@tum.de) are with Molecular Electronics, School of Computation, Information and Technology, Technical University of Munich, Hans-Piloty Straße 1, 85748 Garching, Germany. Marc Tornow is also with the Fraunhofer Institute for Electronic Microsystems and Solid State Technologies (EMFT), Hansastraße 27d, 80686 Munich, Germany.

Feng Liu and Dario Leister are with the Faculty of Biology, Ludwig-Maximilians University, Großhaderner Str. 2-4, 82152, Planegg-Martinsried, Germany.

Thomas Dittrich is with the Helmholtz Zentrum Berlin für Materialien und Energie GmbH, Schwarzschildstr. 8, Berlin, Germany.

Jerry Fereiro is with the Indian Institute of Science Education and Research, Thiruvananthapuram, School of Chemistry, Kerala 695551, India.

David Cahen (david.cahen@weizmann.ac.il) is with the Weizmann Institute of Science, Rehovot, 76100, Israel.

however, rely on complicated chemistry and cumbersome processing of the bottom gold contact. Nevertheless, although gold as the bottom substrate has been extensively explored, other material alternatives could provide additional flexibility and compatibility with the established microelectronics industry. That is the case for titanium nitride (TiN), a hard, conductive ceramic, well-established in CMOS technology, chemically inert, and biocompatible. In contrast to gold, which must be post-processed to produce an ultra-flat surface (template-stripped gold), TiN can present a low roughness (less than a nanometer) as prepared. Moreover, TiN surfaces can also be functionalized with well-defined organophosphonate SAMs [17-19].

The present paper presents a readily implemented way to functionalize the TiN surface by using mixed SAMs to direct the orientation of PSI monolayers. Moreover, by using molecules with a hydrophilic group at the end of the alkane chain, the immobilization of proteins can be successfully achieved. Figure 1 b) shows the SAM molecular structure as an inset. The side group R can be amine ($NH_2$), carboxyl (COOH) or hydroxyl (OH). By choosing a combination of R groups to form a determined mixed SAM, in our case OH-$NH_2$ and OH-COOH, we could modulate the electrostatic charge of the SAM, positive and negative, respectively, when in contact with the protein solution. It has been pointed out that mixed monolayers can modulate the work function of the functionalized substrates and may offer more stable SAMs [20]; this is particularly important for COOH termination, which is challenging to produce in a well-organized manner. A PyMOL calculation of the electrostatic surface charge distribution on the PSI trimer surface at pH=8.0 suggests that negative and positive charges are asymmetrically distributed [6]. The stromal side, where the sulfide-iron redox center is located, is mainly positively charged, and the lumen side, where the P700 can be found, is predominantly negatively charged. This supports our assumption that differently charged SAMs might guide the PSI orientation during incubation for the formation of a self-assembled monolayer.

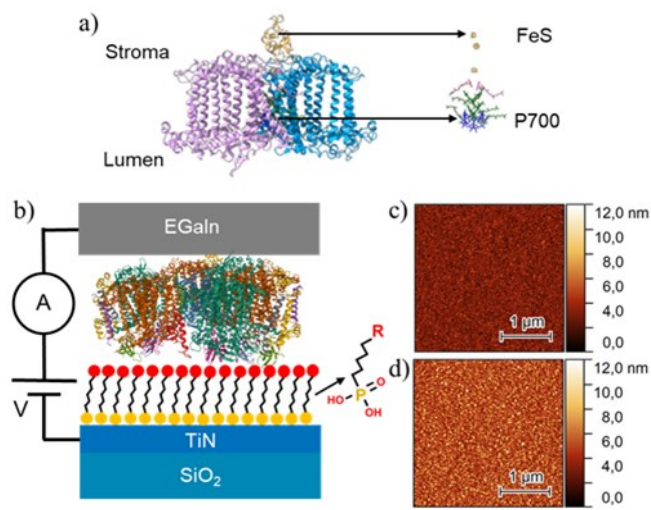

**Figure 1.** a) Left: PSI core (retrieved form protein data bank, PBD : 5OY0) Polymeric subunits PsaA (Pink), PsaB (Blue) and PsaC (Yellow), Right: Electron transfer chain from bottom to up: P700 special pair (purple), other two pairs of chlorophyll (green), two phylloquinones (pink) and three iron-sulfide (FeS) clusters. Black arrows indicate the position of the electron chain inside PSI. b) Directed PSI trimer (PDB:5OY0) immobilization scheme on a self-assembled phosphonate molecules monolayer on TiN with different residual groups R: hydroxyl, amine or carboxyl (dimensions drawn are exaggerated for visualization purposes). The electrical characterization circuit scheme indicates that the bottom substrate, TiN, is grounded, and the top EGaIn contact is biased. The voltage is applied and the current is measured with a source measure unit. AFM tapping mode images of TiN and of PSI on TiN / SAM, (c) and (d), respectively.

Electrical measurements using a eutectic gallium indium (EGaIn) soft metallic top contact indicate a preferential orientation of the protein, manifested by an asymmetrical behavior of the current density if the bias (applied to the EGaIn) is either negative or positive. It has been suggested that the molecular dipole orientation in a given junction can introduce rectification properties [2, 21, 22]. Likewise, such molecular dipoles would impact the electronic properties of the surface and, among other effects, are expected to shift its work function [21, 23-26].

TABLE I    AFM characterization: roughness, protein coverage and thickness (errors are standard deviations)

| Sample | RMS roughness (nm) | Protein Coverage (%) | Thickness (nm) |
|---|---|---|---|
| TiN | 0.7 ± 0.1 | - | |
| OH/COOH/NH2 SAM | 0.7 ± 0.1 | - | 0.8 ± 0.6 |
| PSI on TiN | 1.8 ± 0.1 | 80 ± 15 | 5.8 ± 2.0 |
| PSI on OH-SAM | 1.4 ± 0.1 | 90 ± 5 | 4.7 ± 1.6 |
| PSI on OH-COOH SAM | 1.1 ± 0.1 | 85 ± 15 | 6.0 ± 1.7 |
| PSI on OH-NH2 SAM | 1.1 ± 0.1 | 85 ± 15 | 5.8 ± 1.9 |

We measured the samples' work function to test the hypothesis that PSI's internal electric dipole guides the junction's rectification properties. Work function shifts for the protein and/or SAM samples were observed relative to the TiN reference or the SAM on TiN.

## II. Experiments and Results

### A. Fabrication method

Photosystem I complexes were obtained from *Synechocystis sp*. PCC6803 cells. PSI trimer was purified by anion ion exchange chromatography and sucrose density gradient high-speed centrifugation. The substrate chips (10 x 10 mm$^2$) of 100 nm TiN sputtered on 850 nm SiO$_2$ on Si (100) were obtained from Fraunhofer EMFT, Munich, Germany. Their root mean square (RMS) roughness (measured by atomic force microscopy in tapping mode) was about 0.7 nm. The samples were cleaned in an ultra-sonic bath sequentially in acetone, isopropanol, and DI water. After they were dried with a stream of nitrogen, they were exposed for 30 sec to oxygen plasma (pressure: 0.3 mbar, power: 100W). Mixed SAM grafting was done via immersion for 24 hours in the respective molecule solution in isopropanol. To prepare an OH-COOH SAM (or an OH-NH$_2$), the substrate was immersed in a 2 mM solution (1 mM per molecule) of 6-hydroxyhexylphosphonic acid and 6-phosphonohexanoic acid (or 6-aminohexylphosphonic acid).

After immersion, the samples were gently rinsed with fresh isopropanol and annealed for 1 hour at 130°C. A short sonication, no longer than 30 seconds, was done to remove only physisorbed molecules on the surface. Protein incubation was done by wetting the substrate with a few drops of protein solution over the substrate until it was completely covered. The substrate with the drop casted protein solution was then stored in a sealed and humid box inside a fridge (4°C) for a week. Long incubation periods are necessary to compensate for a low protein concentration solution (0.5 to 1 μg/μL). An additional approach was tested by incubating proteins on bare TiN, previously exposed to O$_2$ plasma (without any linker monolayer).

The samples were prepared for electrical characterization with EGaIn by gluing them with silver paste on a copper sheet. The TiN surface was shorted electrically with the copper sheet by scratching one corner of the sample top surface and connecting it to the copper with silver paste.

### B. Topographical and thickness characterization with atomic force microscopy

Dedicated samples were fabricated to be characterized with the help of atomic force microscopy (AFM). TiN samples with SAM and SAM / protein layers were scanned in AFM tapping mode. The thickness of the molecules and protein layer can be accurately determined if the sample's surface is scratched on a 1 x 1 μm$^2$ region with the AFM tip (AFM in contact mode). Then, a tapping mode image of 3 x 3 μm$^2$ size is taken at the same position. From such an image, a histogram of heights that feature two peaks can be extracted. Those correspond to the level height of the scratched and non-scratched areas. The thickness of the monolayer is the difference between both peaks.

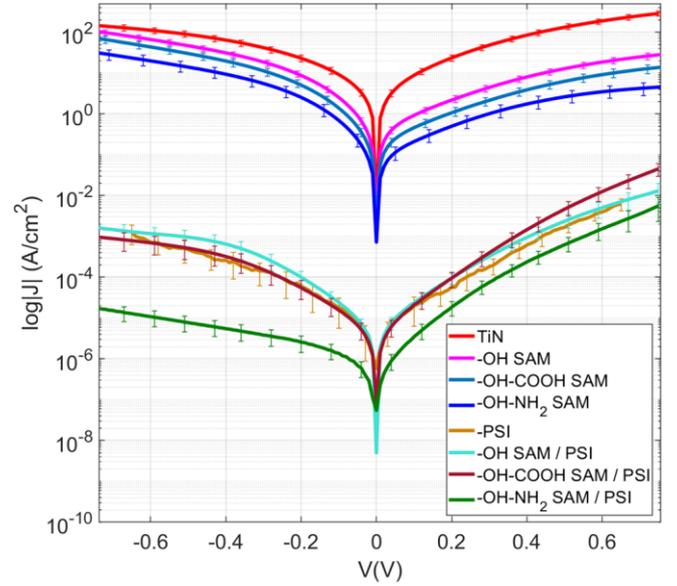

**Figure 2.** Voltage-current density characteristics for TiN/mixed SAM/PSI and TiN/PSI (Brown)

In Table I, the values of the thickness for SAM and protein layers are summarized. In Figures 1 c) and d), typical tapping mode images of TiN (c) and PSI on TiN/OH-COOH SAM (d) can be seen. A regular pattern of globular nanostructures can be spotted in Figure 1 d) which cannot be seen in Figure 1 c). Those objects have a size of around ~50 nm. Nominally, PSI trimers have a diameter of around 20 nm. Therefore, considering the resolution limit of the AFM tip (20 nm diameter), we assign these objects to the protein complexes.

RMS roughness information can be extracted from the tapping mode images. Table I summarizes the samples' RMS roughness, coverage, and thickness. The coverage was calculated by using the flooding function of the AFM analysis software WSxM 5.0. The flooding threshold was each sample's average thickness value. The coverage percentage corresponds to the surface with protein, excluding the scratched "hole" area: The protein monolayer increases the roughness of the surface, and its thickness agrees well with values found in the literature for PSI monolayers in a dry state on gold or SiO$_2$ [27]. PSI is a membrane protein with a height of around 9 nm; as a PSI thickness of ~6 nm is observed, the protein may have shrunk, probably due to dehydration. The estimated distance between the P700 reaction center and the iron-sulfide cluster (electron transfer chain) is ~6 nm (PDB:5oy0) [28, 29]. Therefore, the measured thickness may correspond to the protein electron path length. Protein conformation and activity is also confirmed by UV-VIS spectra taken from immobilized proteins on TiN, which showed absorption peaks at about 439 and 671 nm, the same peaks found in spectra of the protein in solution (data not shown).

### C. Electrical characterization

Figure 1 b) shows the electrical measurement scheme. The sample's bottom contact is grounded, and the top EGaIn soft contact is biased. A source measure unit (SMU) applies voltage and measures current. In the setup, the EGaIn cone hangs from a syringe that serves as a container. Micromanipulators allow the sample to move in the x, y, and z directions.

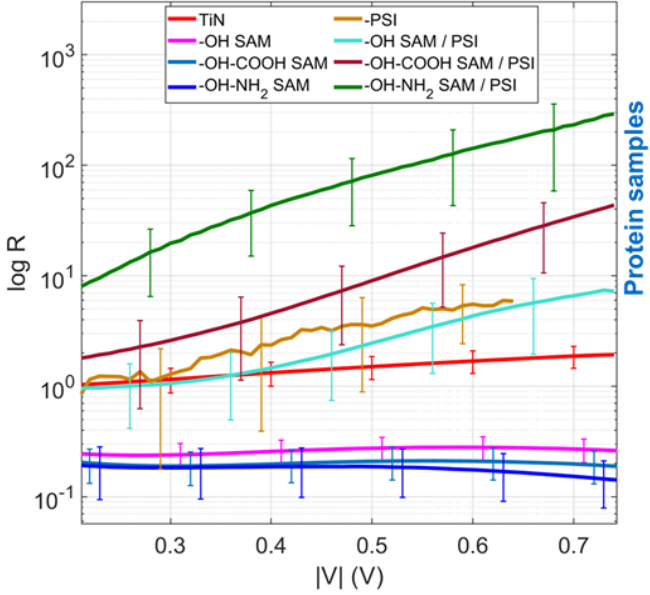

**Figure 3.** Log rectification ratio vs. absolute voltage for TiN/mixed SAM/PSI. Protein samples featured R>1, and SAM samples featured R<1

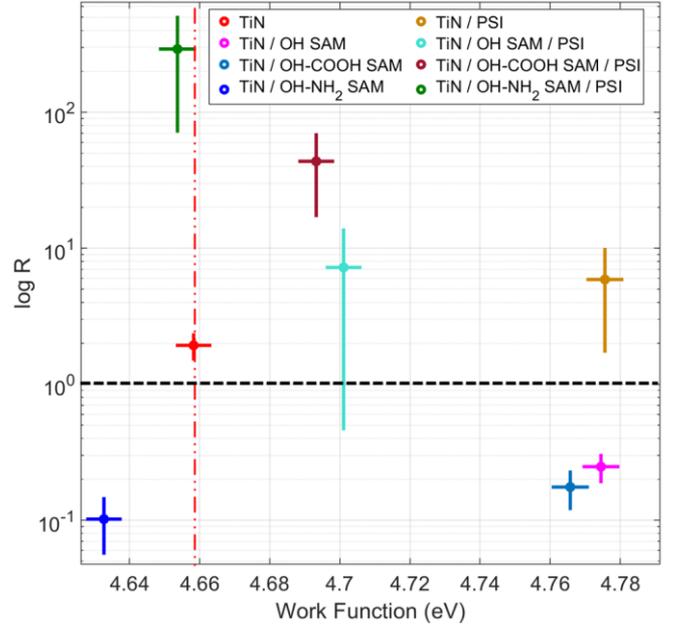

**Figure 4.** Log Rectification ratio at maximum voltage vs. work function values for different samples. The red dashed-dotted line indicates the TiN work function. Samples with protein have R>1 (above the dashed black line), and samples with SAM have R<1 (below the dashed black line).

The syringe with a hanging EGaIn drop is at first attached to a cleaned gold surface; by carefully stretching out the EGaIn from the surface, a cone can be created, and with the help of a calibrated camera, the diameter is fixed to 70 μm when it lands on the sample. Once the junction is formed, several cycles sweep from 0 to negative or positive voltage bias. Ten junctions were created and measured per sample to decrease the measurement uncertainty arising from the manual cone formation.

Figure 2 plot shows the absolute current density $|J|$ (A/cm$^2$) in logarithmic scale as a function of the applied voltage for several samples, including reference bare TiN, TiN/SAM, and PSI on TiN/SAM samples. The error bars indicate the 80% confidence interval of 10 measurements on the same sample, where each measurement is the complete second sweep from negative to positive bias. This bias sweep was selected after a systematic study with the EGaIn setup that suggested that the first sweep and complete cycle were more prone to impact the rectification ratio. Overall, the TiN bare substrate has the highest current density, followed by the TiN/SAM samples, and PSI samples feature the lowest current densities.

By comparing the current densities at maximum voltages, it can be seen that the protein junctions' current density at positive bias is higher than at negative bias by at least one order of magnitude. This behavior is qualitatively the opposite for TiN/SAM samples, where the current density is higher at negative bias than at positive. Such characteristics can be quantified by defining the so-called rectification ratio (R):

$$R = \frac{|J(+)|}{|J(-)|} \quad (1)$$

where $|J(+/-)|$ is the current density at positive or negative bias, respectively. Figure 3 shows log R as a function of the absolute voltage. As can be seen, R > 1 for samples with protein, R < 1 for mixed SAM samples and R~1 for bare TiN references.

At biases of +/- 0.75V, a relatively large rectification number was found for PSI on OH-COOH and OH-NH2 SAMs (~40-200). These rectification ratio numbers for (bio-) molecular junctions are in agreement with previously reported results [2, 12]. The direction of current rectification suggests that the protein might have adopted a "down" orientation, with the P700 redox center near the EGaIn top contact. In such a position, the internal electric dipole is aligned in the direction "up". It is worth pointing out that electrical measurements of PSI on the TiN hydrophilic surface ($O_2$ plasma exposed) without SAM were also possible; the electrical measurements however, were unstable, and electrical shorts emerged at a high bias (>0.65V). This can be observed in Figures 2 and 3, where the average IV and log R curves of the TiN-PSI sample are noisier than the others.

*D. Kelvin probe measurements*

Samples were investigated with a Kelvin probe setup (Besocke) using a gold grid as the reference electrode. The measured parameter is the contact potential difference (CPD) between the reference electrode work function (WF) and the sample WF. From an initial calibration measurement concerning a highly oriented pyrolytic graphite sample, HOPG (WF=4.475 eV) [30], the work function of the gold reference electrode was determined to be 4.755±0.005 eV.

Figure 4 shows the log R at maximum bias vs. work function plot for equivalent samples to those reported in Figure 2. The dashed black line indicates the limit between R > 1 (above) and R < 1 (below). The dashed-dotted red line indicates the TiN reference WF. OH and OH-COOH SAMs shifted towards higher WF values relative to the TiN WF. On the contrary, OH-NH$_2$ SAM shifted to a lower WF. Regarding protein samples, a substantial shift towards higher WF was observed for PSI on TiN (without SAM), followed by PSI on OH and OH-COOH SAMs. The sample with PSI on OH-NH$_2$ SAM shifted to lower

WF values. It was suggested in the introduction that electric dipoles should induce changes in the WF of the surfaces depending on their orientation. Hence, the hypothesis to evaluate here is to determine if there is a correlation between the rectification ratio and the work function shift, which may indicate a possible orientation of the dipoles on the surface. A dipole pointing upwards (downwards) from the surface would induce a lower (higher) WF because the extraction of electrons from the surface would be energetically more favorable (less favorable) and, at positive (negative) bias, could induce $R > 1$ ($R < 1$). As can be seen in Figure 4, such reasoning only is valid for the sample with PSI on OH-NH$_2$, which features $R > 1$ and WF shifts to lower values, and for samples with OH, OH-COOH SAMs which feature $R < 1$ and WF shifts to higher values. The expected trend between R and WF shifts is not present for the other samples: PSI on TiN/OH-COOH SAM and OH-NH$_2$ SAM Therefore, a direct correlation between WF shifts and rectification ratios seems too general, and the WF shifts should be analyzed by considering only similar linker SAM samples. In such a situation, relative WF shifts allow us to speculate a preferential orientation if it is towards higher or lower values for PSI immobilized on OH-COOH SAM (negatively charged) or OH-NH$_2$ SAM (positively charged). Hence, in the first case (PSI on OH-COOH), the protein may have adopted a down orientation, and for the second scenario (PSI on OH-NH$_2$), the protein may have adopted an up orientation. Regarding dipole orientation, they may align upwards (PSI on OH-COOH) or downwards (PSI on NH$_2$-COOH). That hypothesis is not confirmed by the electrical measurements as $R > 1$ for all PSI samples. Empirically, PSI tends to immobilize on hydrophilic surfaces in a preferred orientation, and flipping the ensemble of proteins involves a dynamic process that is yet to be understood.

III. DISCUSSION AND CONCLUSION

The method reported in this paper demonstrates a bottom-up technique that orients PSI ensembles on ultra-flat sputtered TiN with the simple use of self-assembled organophosphonate molecules, which are widely commercially available. The relatively high rectification ratios achieved with our technique have the same order of magnitude as other studies using C60 cages. The idea of protein internal dipole as the origin of the rectification was explored using Kelvin probe contact potential difference measurements. The results indicated a shift in the TiN work function for all samples; in particular, PSI samples' WFs shifted in different directions (the relative change of the WF had different signs), depending on the underlying SAM. As the junctions are complex, more experiments must be done to understand the protein dipole's orientation with respect to the underlying SAM, which may also feature a dipole moment itself. In particular, a molecular simulation of the underlying SAM should provide information about the dipole moment of the molecules. Another potential experiment might take advantage of PSI being optically active; therefore, future work using surface photovoltage measurements will evaluate such inquiry. The role of the linker layers can also be investigated using impedance spectroscopy methods to determine the contact resistance of the junctions. Another potential experiment is chemical peptide conjugation using EDC-NHS to covalently bind the OH-COOH linker SAM with PSI. Work in progress also involves the fabrication of semi-transparent solid-state top contacts for wide-range temperature-dependent photo-electrical measurements.

ACKNOWLEDGMENT

We gratefully acknowledge technical support by Rosemarie Mittermeier, sample processing support by Benedikt Schoof and data analysis discussions with Christian Pfeiffer.